# An introduction to the Patstat database with example queries

Gaétan de Rassenfosse [a], Hélène Dernis [b], Geert Boedt [c]

[a] The University of Melbourne, MIAESR and IPRIA, Level 5, 111 Barry St., Carlton VIC 3010, Australia. Email: gaetand@unimelb.edu.au. *Corresponding author*.
[b] The Organisation for Economic Co-operation and Development, 2 rue André Pascal, 75775 Paris Cedex 16, France.
[c] European Patent Office, 12 Rennweg, 1030 Vienna, Austria.

This version: April 1, 2014

**Abstract**
This paper provides an introduction to the Patstat patent database. It offers guided examples of ten popular queries that are relevant for research purposes and that cover the most important data tables. It is targeted at academic researchers and practitioners willing to learn the basics of the database.

1. **Introduction**

Empirical research on the economics and management of innovation is benefiting from greater availability of structured data. The most prominent database is certainly the European Patent Office (EPO) Worldwide Patent Statistical Database, henceforth 'Patstat'. Patstat offers bibliographic patent data for more than 100 patent offices, sometimes as early as the nineteenth century. It is a valuable tool for the community of researchers because it contains raw data that is collected in a transparent manner. This rich database promises to greatly improve the quality of empirical research in the field. It is, however, difficult to navigate in the wealth of data it offers and many prospective users are deterred by its apparent complexity.

This document seeks to demystify Patstat and offers guided examples on a broad range of queries.[1] It is assumed that the reader has a general knowledge of Structured Query Language (SQL).[2] We have used the April 2013 edition of the database and rely on the MySQL language. Users of another dialect of SQL may have to slightly adapt the queries. Our guiding philosophy in creating the queries was to cover the most important tables and to exploit useful SQL commands. We devote particular attention to outlying some potential uses of the queries for research purposes as well as explaining pitfalls of the data. The reader shall refer to the 2009 OECD Patent Statistics Manual for further guidelines for building and interpreting patent data.

---

[1] This document focuses on the offline Patstat database, available in a series of DVDs from the EPO. A specific introduction exists for the online version of Patstat ("Sample Queries and Tips – Patstat online", available on the EPO website). The online version offers visualisation tools and linked resources, but is less flexible than the offline version.
[2] In particular, we assume knowledge of joins, groups, views and embedded queries. Many introductory courses to SQL are freely available online, including one on the EPO website.

In a desire to make this introduction accessible to the greatest number, we have produced a test database in MS Access format. This database contains the relevant data used in this paper as well as all the queries. It allows readers to familiarise themselves with the Patstat database without having to install it on their computers. The test database and the queries are available at http://www.runmycode.org or upon request from the authors.

## 2. Patstat cookbook

The Patstat database consists in a set of tables that follow a relational database schema where tables can be connected to each other using a relevant entry key.[3] The table on patent applications, labeled *tls201_appln*, contains more than 74 million records and is the central element of Patstat, as indicated in Figure 1. The other tables contain information on each of the patent applications, *e.g.*, inventors and owners, technology fields, titles and abstracts, publication instances, and citations.

**Figure 1.** Patstat database schema

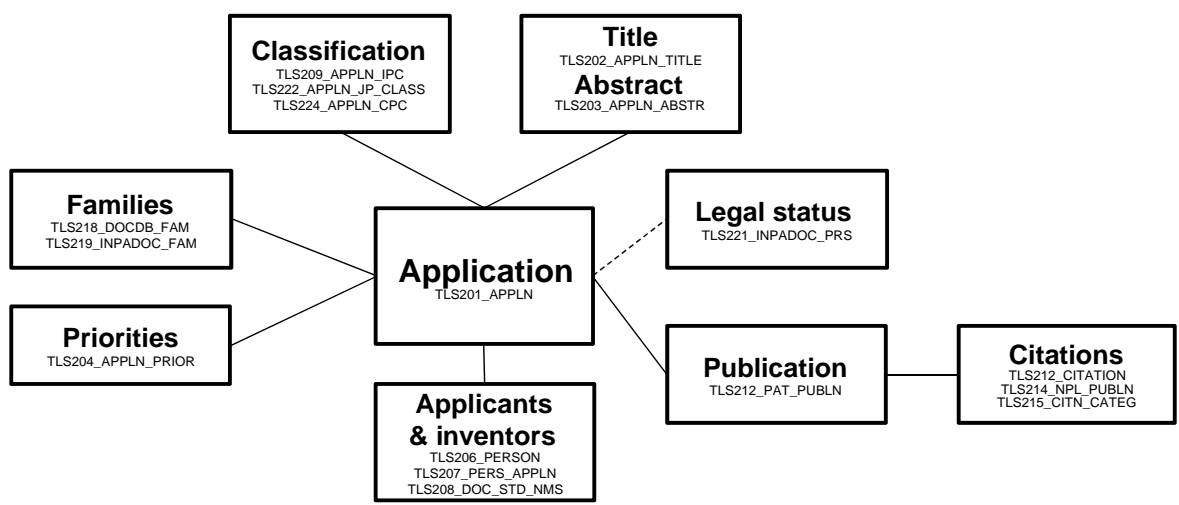

*Source:* European Patent Office, Patstat database, April 2013. Not all the tables are reported.

To limit the volume of data retrieval, we run our queries on a sample of patent applications describing inventions related to wind turbine technologies and filed in the year 2005 anywhere in the world. Patent applications related to wind turbines technologies are predominantly found in the IPC subclass F03D (Dubarić et al. 2012).[4] This subclass includes all the IPC codes that start with F03D such as 'F03D 1/00' (wind motors with rotation axis substantially in wind direction) and 'F03D 5/02' (wind-engaging parts being attached to endless chains or the like).

---

[3] The identifier of patent applications (*appln_id)* is frequently used to link tables with each other. A full description of tables and fields is provided in the Data Catalog, which is available on the Patstat DVDs and can also be downloaded from the EPO website.

[4] IPC stands for International Patent Classification. IPC codes are used by patent examiners to identify the areas of technology to which patents pertain. Note that not all patents have IPC codes. Wind energy patents can also be identified using the CPC code Y02E10/70 available in *table tls224_appln_cpc*. CPC stands for Cooperative Patent Classification and is a joint classification system between the US Patent and Trademark Office (USPTO) and the EPO.

## 2.1 Identification of patents by technology fields

We use the term 'application' to refer to entries in table *tls201_appln*. This table lists all the applications available in the Patstat database and assigns them a unique and stable *appln_id*, which is built from a combination of the patent authority (the patent office where the application was submitted), the patent application number, and the application kind code (indicating for example whether the application is a patent application, a PCT application or a design application). The application number is distinct from the application identifier. The application number is the number issued by the patent authority where the application was filed, whereas the application identifier is specific to the Patstat database. The latter is called a 'primary key' in SQL jargon. In Query 1, the *appln_id* is used to link table *tls201_appln* with table *tls209_appln_ipc*, which contains the IPC codes assigned to each application.

**Query 1**
```sql
SELECT
    DISTINCT t1.appln_id, t1.appln_auth, t1.appln_nr, t1.appln_kind
FROM
    tls201_appln t1
        INNER JOIN
    tls209_appln_ipc t2 ON t1.appln_id = t2.appln_id
WHERE
        year(t1.appln_filing_date) = 2005
        AND (t1.appln_kind = 'A' OR t1.appln_kind = 'W')
        AND t2.ipc_class_symbol LIKE 'F03D%'
ORDER BY t1.appln_auth, t1.appln_id;

CREATE VIEW our_sample AS
[paste above query here];
```

Note: See main text for comment regarding compatibility of the query with MS SQL.

The first statement extracts: the unique application identifier (*appln_id*); the two-letter code of the patent application authority (*appln_auth*); the patent application number (*appln_nr*) and the kind of application (*appln_kind*). We select *patent applications* by choosing applications with an *appln_kind* code of either 'A' (direct filing) or 'W' (PCT application, see section 2.2). The SELECT DISTINCT clause is used to avoid duplicates in the result table in case a given patent application has more than one IPC code starting with F03D. The query returns 2125 distinct patent applications and sorts them by *appln_auth* and *appln_id*. The first five results are presented in Table 1 for illustrative purposes. Note that the use of ORDER BY generally slows down queries and can usually be avoided.

**Table 1.** First five rows of Query 1.

| appln_id | appln_auth | appln_nr | appln_kind |
|---|---|---|---|
| 55286477 | AP | 200603687 | A |
| 55286499 | AP | 200603713 | A |
| 532990 | AR | P050100289 | A |
| 533082 | AR | P050100386 | A |
| 533175 | AR | P050100493 | A |

The two-letter code shown in *appln_auth* column in Table 1 corresponds to the receiving office: 'AP' refers to the African Regional Intellectual Property Organization (ARIPO) and 'AR' to Argentina's National Institute of Industrial Property. The codes follow the WIPO ST.3 format.[5] Exceptionally, some codes in Patstat might not have a correspondence (for instance if an applicant cites a patent document with a non-standard country code).

The second statement creates a temporary table (a 'view' in SQL jargon) that is referred to as *our_sample* and contains the set of patents related to wind turbine technologies defined in the first statement. Views are particularly useful to break down queries into smaller, simpler pieces. Views cannot have indexes, so that they are better suited for small populations. Users of MS SQL should remove the ORDER BY keyword from the first query.

### 2.2 Identifying PCT applications

The Patent Cooperation Treaty (PCT) is an international patent law treaty that provides a unified procedure for filing patent applications to protect inventions in each of its contracting states. A patent application filed under the PCT is called an international application, or PCT application. PCT applications are often associated with inventions of high market potential (van Zeebroeck and van Pottelsberghe 2011) and are being used increasingly by patent applicants (Frietsch et al. 2013).[6] Researchers sometimes use them to study the international dimension of patenting activity (see *e.g.* Allred and Park 2007). A PCT application does not automatically lead to global patent protection. Instead, patent applicants eventually need to 'apply' for patents in each of the jurisdictions where they wish to pursue patent protection by starting the national search and/or examination process. These 'national' patents are formally referred to as national-phase entry of PCT applications.

In Patstat, PCT applications at international phase can be identified in different ways. They are associated with an *appln_kind code* 'W' in table *tls201_appln*, and they are associated with a publishing patent authority (*publn_auth*) set to 'WO' in table *tls211_pat_publn*. The two-letter code 'WO' stands for the World Intellectual Property Office (WIPO). National-phase entry of PCT applications can be identified with the field *internat_appln_id* in table *tls201_appln*, which corresponds to the *appln_id* of the PCT application (the field *internat_appln_id* is set to 0 for applications not originating from a PCT filing).

The query below lists all PCT applications that entered the national phase either at the State Intellectual Property Office of the People's Republic of China (SIPO) or at the Japanese Patent Office (JPO) and for which the Danish Patent and Trademark Office (DKPTO) was the receiving office (that is, application authority) of the initial PCT application.

---

[5] See the WIPO's "Recommended standard on two-letter codes for the representation of states, other entities and intergovernmental organizations" (Standard ST.3) for the exhaustive list of codes, available on the WIPO website.

[6] Note that the link between patent value and PCT status is a priori ambiguous. As Guellec and van Pottelsberghe (2001) and Reitzig (2004) point out, patents applicants may be uncertain about the economic success of the patent's underlying invention and use the PCT route to 'buy' additional decision time. Alternatively, the economic success of the patent's underlying invention may be well established at the date of filing and PCT is used to seek global protection as fast as possible.

```
                                                                                    Query 2
SELECT
    t1.appln_id AS PCT_appln_id,
    t1.appln_auth AS PCT_appln_auth,
    t1.appln_nr AS PCT_appln_nr,
    t1.appln_kind,
    t2.appln_id AS appln_id_sf,
    t2.appln_auth AS appln_auth_sf
FROM
    our_sample t1
        INNER JOIN
    tls201_appln t2 ON t1.appln_id = t2.internat_appln_id
WHERE
    t1.appln_auth = 'DK'
        AND t2.appln_auth IN ('CN', 'JP')
ORDER BY t1.appln_id;
```

The statement relies on patents in *our_sample*. It literally selects all the applications submitted at the SIPO or the JPO that have their *internat_appln_id* equal to the *appln_id* of PCT patent applications in *our_sample* submitted at the DKPTO. The first five results (out of a total of 15 national phase entries) are presented in Table 2.

**Table 2.** First five rows of Query 2.

| PCT_appln_id | PCT_appln_auth | PCT_appln_nr | appln_kind | appln_id_sf | appln_auth_sf |
|---|---|---|---|---|---|
| 15563101 | DK | 2005000031 | W | 8300709 | CN |
| 15563116 | DK | 2005000046 | W | 8300768 | CN |
| 15563118 | DK | 2005000048 | W | 8300756 | CN |
| 15563246 | DK | 2005000181 | W | 8306357 | CN |
| 15563258 | DK | 2005000193 | W | 39635652 | JP |

*2.3 Obtaining information on priority status*

A priority patent application is the first patent application filed to protect an invention. Under the 1883 Paris convention, a priority patent can be filed in other jurisdictions, with the aim of extending the protection to other countries. The subsequent patents are called 'second filings'.

Depending of the research goal, the priority status of patent documents is an important piece of information. First, the count of patent applications described in Query 1 may lead to multiple counts of inventions since it mixes both priority and second filings. Counting unique inventions involves counting only priority filings.[7] de Rassenfosse et al. (2013) explain the details of such a 'worldwide count of priority filings'. The issue of double counting becomes less acute if patents are counted at a single office of reference such as at the EPO.[8] Second, it may be desirable to know the priority status of the patent document in order to avoid potential selection bias, especially when patents are counted at a single office of reference. de Rassenfosse et al. (2014) explain that the single-office count may produce biased econometric estimates of patent production functions.[9] They propose a test based on the priority status of the patent application to detect the presence of selection bias.

---

[7] Note that an alternative approach for counting unique inventions involves counting families of patents. More information on patent families is provided in the next section.
[8] One can often observe priority filings and subsequent second filings at the same patent office. This phenomenon is driven by divisional (or similar) applications. If a priority application was filed at the EPO and a divisional was also filed at the EPO, this divisional would claim priority from the original document and is therefore technically equivalent to a second filing. Such cases can be identified with table *tls216_appln_contn*.

The query below returns the priority status of the patent documents in our set.

**Query 3**
```sql
SELECT DISTINCT
    t1.appln_id,
    (CASE
        WHEN t2.appln_id IS NULL THEN 1
        ELSE 0
    END) AS is_a_pf
FROM
    our_sample t1
        LEFT OUTER JOIN
    tls204_appln_prior t2 ON t1.appln_id = t2.appln_id
ORDER BY t1.appln_id;
```

This statement selects all the *appln_id* from *our_sample* dataset and matches them to *appln_id* provided in table *tls204_appln_prior*, which lists priority patents claimed in second filings. By definition, all patent applications that do not claim a priority are priority filings. Therefore, the column labelled *is_a_pf* takes the value 1 if no match is found. Note that, contrary to previous queries, tables are linked together using the LEFT OUTER JOIN statement. This joint returns all rows from the left table (t1) and adds information from the right hand side table (t2) when a match exists. Note also that second filings may claim more than on priority filing in table *tls204_appln*, hence the use of the DISTINCT clause. Query 3 reports 957 priority applications out of 2125 patent applications originally identified in *our_sample*. The first five records are presented in the table below for illustrative purposes.

**Table 3.** First five rows of Query 3.

| appln_id | is_a_pf |
|---|---|
| 65303 | 0 |
| 133780 | 0 |
| 149552 | 1 |
| 151084 | 0 |
| 151176 | 0 |

*2.4 Computing the patent family size*

A patent family refers to a group of patent applications that are all related to each other by way of one or several common priority filings. Following Putnam (1996), researchers use information on patent families as a proxy for patent value. The validity of this approach was established by Harhoff et al. (2003) who show that family size is correlated with estimates of the value of patent rights from a survey of patent-holders. The family size is an internationally-comparable measure of value and is thus well-suited for studies relying on patent applications filed in different jurisdictions.

---

[9] Patent production functions are used in econometric studies to study the determinants of the number of patents produced by an economic unit such as a firm or a country.

The next query counts the patent family size associated with the applications in *our_sample*. We adopt the 'extended' family definition (INPADOC) which captures all applications that are directly or indirectly linked via priority filings. An alternative approach involves using the DOCDB family, available in table *tls218_docdb_fam*. Various definitions (and hence ways to measure) patent families exist and a good overview is provided in the 2009 OECD Patent Statistics Manual as well as in Martínez (2011).

**Query 4**
```sql
SELECT
    t1.appln_id, COUNT(t3.appln_id) AS family_size
FROM
    our_sample t1
        INNER JOIN
    tls219_inpadoc_fam t2 ON t2.appln_id = t1.appln_id
        INNER JOIN
    tls219_inpadoc_fam t3 ON t3.inpadoc_family_id = t2.inpadoc_family_id
GROUP BY t1.appln_id
ORDER BY t1.appln_id;
```

Notice that Query 4 calls table *tls219_inpadoc_fam* twice, under the aliases *t2* and *t3*: *t2* links each *appln_id* from *our_sample* to its patent family identifier *inpadoc_family_id*, and is in turn linked to *t3* to retrieve and count all family members (*t3.appln_id*) that belong to the same *inpadoc_family_id*. The first five rows are presented below.

**Table 4.** First five rows of Query 4.

| appln_id | family_size |
|---|---|
| 65303 | 9 |
| 133780 | 4 |
| 149552 | 14 |
| 151084 | 13 |
| 151176 | 9 |

Researchers are sometimes interested in the number of jurisdictions that the family covers. For example, the OECD produces an indicator on triadic patent families, which captures patents granted by the US Patent and Trademark Office (USPTO) and filed at the EPO and the Japan Patent Office (JPO) to protect the same set of inventions (Dernis and Khan 2004). de Rassenfosse and van Pottelsberghe (2009) show that triadic patents are a good indicator of countries' research productivity (compared with priority filings, which are affected by variations in the propensity to patent across countries). Information on how to identify triadic patents in Patstat is provided in Appendix A. Another family-based indicator is obtained by simply counting the number of jurisdictions identified in a family – we call it here the 'geographic' family size (see also Squicciarini et al. 2013). Query 4 can be easily adapted to measure the geographic family size, as illustrated in Query 5.

```sql
                                                                                    Query 5
SELECT
    t1.appln_id,
    COUNT(DISTINCT t4.publn_auth) AS geog_family_size
FROM our_sample t1
        INNER JOIN
    tls219_inpadoc_fam t2 ON t2.appln_id = t1.appln_id
        INNER JOIN
    tls219_inpadoc_fam t3 ON t3.inpadoc_family_id = t2.inpadoc_family_id
        INNER JOIN
    tls211_pat_publn t4 ON t4.appln_id = t3.appln_id
WHERE t4.publn_auth != 'WO'
GROUP BY t1.appln_id
ORDER BY t1.appln_id;
```

Compared with Query 4, Query 5 uses information from an additional table, *tls211_pat_publn*, to recover information on the patent offices of destination (publication authorities) of all INPADOC family members, and excludes the PCT publication authority (WO) as it has an international coverage.[10] The first five results are presented in Table 5. A comparison with the results presented in Table 4 suggests that large differences may exist between the two measures of family size. For example, while the family associated with *appln_id* number 65303 has nine members, it covers only four jurisdictions: Germany, members of the European Patent Convention (through the EPO), the United States and China. There are various reasons why the family size may differ from the geographic family size such as procedural reasons (unity of invention requirement or maximum number of independent claims) and patent strategy reasons (*e.g.*, creation of patent thickets).

**Table 5.** First five rows of Query 5.

| appln_id | geog_family_size |
|---|---|
| 65303 | 4 |
| 133780 | 4 |
| 149552 | 12 |
| 151084 | 11 |
| 151176 | 8 |

Note that Query 5 reports the number of distinct patent offices and not the number of distinct countries *per se*. This distinction matters when patents are filed at regional offices such as the ARIPO or the EPO, which cover many jurisdictions. Patents granted by a regional office must be validated in each of the member states where patent protection is sought. As a result, while a patent application at the EPO virtually covers a market of approximately 500 million people, its actual coverage could be much smaller depending on the countries in which the patent was validated. One way of dealing with the issue involves adding information on the number of jurisdictions in which regional applications were validated, after the patent has been granted at the regional office considered. This can be done with Patstat but we will not discuss it as it far exceeds to scope of this paper.[11] Note also that patents listed in *our_sample* may belong to the same family, and further consolidation may be envisaged to control for double counting.

---

[10] Indeed, not excluding the PCT application at international phase inflates the family count by one unit. For example, if the JPO is the receiving office of a PCT application that then enters national phase at the JPO only, not excluding the PCT application at international phase will lead to a family size of 2 instead of 1.

[11] Briefly, the approach for EPO patents would be to use the INPADOC legal status database in addition to Patstat and identify the relevant legal status codes that indicate a validation or renewal fee payment in a designated state. The INPADOC database is available as an add-in table to Patstat, as explained in Appendix A.

### *2.5 Counting patents by country (simple counts vs. fractional counts)*

Patent data provides information on inventors and applicants and it is thus a rich source of information about the structure of technology production. Briefly, the inventor country of residence reflects the country of origin of inventions while the applicant country of residence reflects the ownership of inventions. The 2009 OECD Patent Statistics Manual proposes a comprehensive discussion on the choice of the reference country for building patent counts. Two distinct counting approaches can be applied, in response to specific analytical requirements: *simple* count method versus *fractional* count method. Since a large number of patent applications are due to team work, it is likely that more than one inventor has contributed to the protected invention, located in one or several countries. Similarly, several applicants may co-own a unique patent. The fractional count procedure is used to better reflect the contribution of each country and avoid multiple counts of a same patent in different countries.

The list of inventors (applicants) can be identified using two additional tables: *tls207_pers_appln* lists the correspondence between patent application and inventors (applicants), and *tls206_person* provides details on names and addresses. The *person_id* identifier enables to establish the link between these two tables. Note that not all patent documents listed in *tls201_appln* have an entry in *tls207_pers_appln*.

The script below performs a fractional count of inventors' country of residence for patent applications in *our_sample*. Inventors have a field *invt_seq_nr* greater than 0 in table *tls207_pers_appln*, while applicants have a field *applt_seq_nr* greater than 0.

<div style="text-align: right">Query 6</div>

```sql
SELECT
    person_ctry_code, SUM(tot_in_ctry/tot_in_patent) AS fractional_count
FROM
    (SELECT
        t.appln_id,
        ifnull(t1.person_ctry_code, '') AS person_ctry_code,
        ifnull(t1.tot_in_ctry, 1) AS tot_in_ctry,
        ifnull(t2.tot_in_patent, 1) AS tot_in_patent
    FROM
        our_sample t
            LEFT OUTER JOIN
            --> Accounts for missing inventor references in
            tls207_pers_appln table
        (SELECT
            a.appln_id,
            b.person_ctry_code,
            COUNT(b.person_id) AS tot_in_ctry
        FROM
            tls207_pers_appln a
            INNER JOIN tls206_person b ON a.person_id = b.person_id
        WHERE
            a.invt_seq_nr > 0
        GROUP BY a.appln_id, person_ctry_code
        --> Compiles country-level count of inventors per patent
        ) t1 ON t.appln_id = t1.appln_id
            LEFT OUTER JOIN
        (SELECT
            appln_id, MAX(invt_seq_nr) AS tot_in_patent
        FROM
            tls207_pers_appln
        GROUP BY appln_id HAVING MAX(invt_seq_nr) > 0
        --> Compiles total count of inventors per patent
        ) t2 ON t.appln_id = t2.appln_id
    ) our_sample_with_country
GROUP BY person_ctry_code
ORDER BY SUM(tot_in_ctry/tot_in_patent) DESC;
```

Note: See main text for comment regarding compatibility of the query with MS SQL.

The above script is more advanced than previous scripts as it is composed of embedded queries providing intermediary counts for facilitating fractional counts by country. (It is possible to break it into smaller statements using VIEWS) The aggregated counts by country are performed on a selection of fields (named *our_sample_with_country*) extracted using *our_sample* table and two sub-queries. Sub-query *t1* reports the counts of inventors by country and by patent and sub-query *t2* reports the total number of inventors by patent. Output from *t1* and *t2* is then linked to patents in *our_sample* using a LEFT OUTER JOIN statement to account for missing records in *tls207_pers_appln* table. MySQL function 'ifnull()' replaces the missing records by an empty record, and sets the count to 1 where missing (either because the *appln_id* was not found in table *tls207_pers_appln* or because no *person_id* was identified for *invt_seq_nr* greater than 0). Users of MS SQL should use the 'isnull()' function instead. They should also specify that the final count needs to be a float by using [CONVERT(float, COUNT(b.person_id)) AS tot_in_ctry] in query *t1*. Previews of results for the field selection (*our_sample_with_country*) and the final count are presented in Table 6 and Table 7 respectively.

Table 6 shows that all the inventors of *appln_id* 263066 are German. By contrast, one-fourth of inventors of *appln_id* 273390 is Swiss and three-fourth is German. Grouping all the shares by *person_ctry_code* leads to the results presented in Table 7. Among the 2125 patent application in our_sample, 609.5 have not been allocated to a country, and 357.2 patents were due to German inventors. A methodology for recovering missing country codes is presented in de Rassenfosse et al. (2013).

**Table 6.** Five randomly-selected rows of field selection of Query 6 (join from t1 and t2).

| appln_id | person_ctry_code | tot_in_ctry | tot_in_patent |
|---|---|---|---|
| 263066 | DE | 2 | 2 |
| 273390 | CH | 1 | 4 |
| 273390 | DE | 3 | 4 |
| 273768 | JP | 1 | 1 |
| 273769 | JP | 1 | 1 |

**Table 7.** First five rows of Query 6 (fractional count by applicant country).

| person_ctry_code | fractional_count |
|---|---|
|  | 609.5 |
| DE | 357.2 |
| US | 248.0 |
| CN | 155.8 |
| DK | 113.5 |

It is straightforward to adapt Query 6 to applicants' country of residence (using the *applt_seq_nr* field instead of *invt_seq_nr* in *tls207_pers_appln*). It is important to stress that applicant and inventor information provided in Patstat and linked via the *tls207_pers_appln* table corresponds to the information available in the last publication associated with an application. For example, if an EP-B1 publication has different applicant names than the original EP-A1 publication, then only the person information for the B1 publication will be available.[12] To recover the current information, it is possible to link Patstat data with data provided by national patent offices, as explained in Section 2.9.

### *2.6 Identifying patents resulting from international collaborations*

The information on applicants and inventors has been used to study among other questions international R&D collaboration (Guellec and van Pottelsberghe 2001; Picci 2010; Danguy 2014), R&D offshoring (Thomson 2013), or network of inventors (Balconi et al. 2004; Ejermo and Karlsson 2006). To the best of our knowledge, only a limited number of studies assess the validity of these indicators. One such study is Bergek and Bruzelius (2010), which casts some doubt on the use of inventor data to measure R&D collaboration.

An example of query identifying patents resulting from international collaboration is presented below. The rationale being that patent applications for which the field *nb_locations* is greater than 1 involve inventors that reside in different countries and are thus the outcome of an international collaboration (i.e., co-invented patents).

---

[12] Persons are also linkable to publications since the October 2013 release of Patstat. Kind code 'A1' refers to a European patent application published with European search report, and 'B1' to a European patent granted.

```sql
                                                                                      Query 7
SELECT
    t1.appln_id, COUNT(DISTINCT t3.person_ctry_code) AS nb_locations
FROM
    our_sample t1
        INNER JOIN
    tls207_pers_appln t2 ON t1.appln_id = t2.appln_id
        INNER JOIN
    tls206_person t3 ON t2.person_id = t3.person_id
WHERE
    t3.person_ctry_code IS NOT NULL
        AND t2.invt_seq_nr > 0
GROUP BY t1.appln_id
ORDER BY COUNT(DISTINCT t3.person_ctry_code) DESC, t1.appln_id ASC;
```

Query 7 counts the number of distinct inventor countries listed in each patent application in *our_sample*. It reports a positive international collaboration conditional on the availability of records in *tls207_pers_appln* table or in the *person_ctry_code* field in table *tls206_person.* The first five results are presented in Table 8.

**Table 8.** First five rows of Query 7.

| appln_id | nb_locations |
|---|---|
| 48145305 | 3 |
| 273390 | 2 |
| 4975233 | 2 |
| 4979189 | 2 |
| 5804835 | 2 |

*2.7 Counting citations received*

Following early works by Carpenter et al. (1981) and Trajtenberg (1990) citation data is used as an indicator of quality – broadly defined as the technological merit and the economic potential of an invention. Note that other indicators of patent quality exist, see in particular the recent work by Squicciarini et al. (2013). Citation data is also frequently used to track knowledge flows (Jaffe et al. 1993) and to measure the speed of knowledge obsolescence (Clark 1976; Jaffe and Trajtenberg 1996). While patent citation data may offer very rich insights, it must be used with caution. One must pay close attention to the effects of the institutional environment on the relevance of citation data as an economic indicator. In particular, patent citations practices differ across patent offices (Michel and Bettels 2001), and examiner-added citations may add extra noise to the data (see Alcácer and Gittelman 2006 for USPTO evidence). In addition, many publications from different patenting authorities but covering the same invention can be cited, leading to a fragmentation of citation records as explained in Webb et al. (2005).

The next query counts the number of citations received in a three-year time window from patent applications published by the German Patent and Trade Mark Office by patent applications published by the EPO. Citations received by a patent are often referred to as 'forward' citations, in opposition to 'backward' citations, which indicate citations made by a patent. The latter is sometimes also called 'references' (by analogy to the reference list of a scientific paper).

**Query 8**

```sql
SELECT
    t1.appln_id, COUNT(distinct t3.pat_publn_id) AS cites_3y
FROM
    our_sample t1
        INNER JOIN
    (SELECT
        appln_id, MIN(publn_date) AS earliest_date
    FROM
        tls211_pat_publn
    GROUP BY appln_id) t2 ON t1.appln_id = t2.appln_id
        INNER JOIN
    tls211_pat_publn t2b ON t2b.appln_id = t2.appln_id
        INNER JOIN
    tls212_citation t3 ON t2b.pat_publn_id = t3.cited_pat_publn_id
        INNER JOIN
    tls211_pat_publn t4 ON t3.pat_publn_id = t4.pat_publn_id
WHERE
    t2b.publn_auth = 'DE'
        AND t4.publn_auth = 'EP'
        AND YEAR(t2.earliest_date)!= 9999
        AND YEAR(t4.publn_date)!= 9999
        AND t4.publn_date <= DATE_ADD(t2.earliest_date, INTERVAL 3 YEAR)
GROUP BY t1.appln_id
ORDER BY COUNT(distinct t3.pat_publn_id) DESC, t1.appln_id ASC;
```

Note: See main text for comment regarding compatibility of the query with MS SQL.

Note that the citation records are based on the published patent documents, hence the use *publn_auth* from table *tls211_pat_publn* instead of *appln_auth* from table *tls201_appln*. The field *publn_auth* captures the publication authority of the patent document. The publication authority is often also the receiving office (*appln_auth*) except in the case of PCT applications where the publication authority is WIPO and the receiving office is the patent office where the patent application was actually lodged. Thus, an alternative to criterion [t2.publn_auth = 'DE'] is [t1.appln_auth = 'DE' AND t1.appln_kind = 'A']. The use of a time window is important when working with patents of different age cohorts in order to avoid data truncation. It is easily implemented with the function 'DATE_ADD()'. Users of MS SQL should use [DATEDIFF(YEAR, t2.earliest_date, t4.publn_date) <= 3] instead. In order to better estimate the citation lag, the date of reference is set to the earliest date of publication of the cited patent. Note that the count is fairly naïve for reasons explained above as well as because it does not take into account the type of EPO citations. See Webb et al. (2005:8) for an overview of citation types at the EPO. The first five results are presented in Table 9.

**Table 9.** First five rows of Query 8.

| appln_id | cites_3y |
|---|---|
| 14995919 | 5 |
| 14997816 | 2 |
| 14971868 | 1 |
| 14974947 | 1 |
| 14975309 | 1 |

*2.8 Obtaining grant information*

A published patent application provides legal rights and some economic benefits to its owner (see *e.g.* Guellec et al. 2012), but most of the value of a patent is achieved when the patent is granted and the owner can enforce its exclusive right. The grant status is therefore an important economic variable. Query 9 shows how to recover information on whether patent applications in *our_sample* filed at the UK Intellectual Property Office (UKIPO) have been granted.

**Query 9**
```sql
SELECT
    t1.appln_id, MAX(t2.publn_first_grant) AS granted
FROM
    our_sample t1
        INNER JOIN
    tls211_pat_publn t2 ON t1.appln_id = t2.appln_id
WHERE t1.appln_auth = 'GB'
        AND t1.appln_kind = 'A'
GROUP BY t1.appln_id
ORDER BY t1.appln_id;
```

The query uses information from table *tls211_pat_publn*. Each application is associated with one or more published documents, and each published document is tagged with an office-specific publication kind code to indicate the kind of publication. The Patstat team has identified the publication kind codes associated with granted documents and the earliest document of an application corresponding to a grant is given a value of 1 in the field *publn_first_grant*. All other documents are given a value of 0. A simple way of finding whether a patent application was granted is thus to select the maximum value of the field *publn_first_grant* for each *appln_id*. If the maximum value is 1, the patent was granted. The status of a patent application associated with a value of 0 is unclear: other legal statuses include, but are not limited to: pending; withdrawn; and refused. Detailed information on legal status can be recovered from table *tls221_inpadoc_prs* for some patent offices (see Appendix A for details). For other offices, it is necessary to link Patstat data with data provided by national patent offices, as explained in the next section. Note that PCT applications are never granted *per se*; only applications that entered the national phase can be granted.

**Table 10.** First five rows of Query 9.

| appln_id | granted |
|---|---|
| 21465239 | 1 |
| 21466952 | 0 |
| 21467768 | 0 |
| 21470294 | 0 |
| 21471154 | 0 |

## *2.9 Linking Patstat with data provided by national patent offices*

It is sometimes desirable to enrich Patstat data with data directly provided by national patent offices, for example to get accurate information on the legal statuses of patent applications or to collect information on reassignments. This can be done by using information from the field *publn_nr* in table *tls211_pat_publn*. The reconstruction of the publication number is specific to each patent office and Query 10 focuses on the rather simple example of the UKIPO.

**Query 10**
```sql
SELECT DISTINCT
    t1.appln_id,
    t2.publn_nr AS publn_nr_patstat,
    CONCAT('GB',RIGHT(t2.publn_nr,7)) AS publn_nr_ukipo
FROM
    our_sample t1
        INNER JOIN
    tls211_pat_publn t2 ON t1.appln_id = t2.appln_id
WHERE t1.appln_auth = 'GB'
        AND t1.appln_kind = 'A'
        AND t2.publn_kind != 'D0'
ORDER BY t1.appln_id;
```
Note: See main text for comment regarding compatibility of the query with MS SQL.

The online patent document and information service of the UKIPO (Ipsum) requires the publication number to be in the following format: 'GBnnnnnnn', *i.e.* the characters 'GB' followed by 7 digits. Query 10 thus appends the characters 'GB' in front of the last 7 digits of the field *publn_nr* in order to recompose a publication number that is compatible with the UKIPO online service. Users of MS SQL need to replace the last element of the SELECT DISTINCT clause with [('GB'+RIGHT(t2.publn_nr,7)) AS publn_nr_ukipo] Note the exclusion of documents with *publn_kind* code 'D0' which, for the UKIPO, correspond to patent applications filed. The field *publn_nr_ukipo* can now be used to search for additional information on the UKIPO website. More generally, one must reverse-engineer the Patstat format to the format use with the national patent office.

**Table 11.** First five rows of Query 10.

| appln_id | publn_nr_patstat | publn_nr_ukipo |
|---|---|---|
| 21465239 | 2410379 | GB2410379 |
| 21467768 | 2423650 | GB2423650 |
| 21470294 | 2441770 | GB2441770 |
| 21471154 | 2424926 | GB2424926 |
| 21471862 | 2425334 | GB2425334 |

## 3. Concluding remarks

This paper has provided a broad overview of the Patstat database by discussing typical queries that rely on the main tables. A good way to proceed from here is to slightly alter the queries and observe how result-sets returned are affected. We hope that users will be able to devise indicators tailored to their research needs and therefore contribute to further improving the quality of empirical research in the fields of economics and management of innovation. In order to avoid duplication of work, however, we encourage researchers to share their contributions with the broad community. Appendix A briefly describes add-ons provided by institutions or individual contributors to enrich Patstat data.

A large community of users has emerged over time and is keen to share its experience and answers questions of beginners on the Patstat forum on the EPO website. An additional helpful resource is the annual Patent Statistics for Decision Makers conference (and the preceding user workshop) where the Patstat community gathers and exchanges recent developments.

**Acknowledgements**

The authors are grateful to Monica Coffano, Jérôme Danguy, Paul Jensen, Catalina Martínez, Clinton McCarthur, Nico Rasters and Gianluca Tarasconi for helpful comments. The paper has also benefited from comments by participants at a staff development workshop at IP Australia (Canberra).

**Appendix A. Resources for Patstat**

*EPO Worldwide Legal Status database* – Also known as *tls221_inpadoc_prs* table, it contains information on legal events that occurred during the life of a patent, either before or after grant. Typical events are: payment of (national) renewal fees, lapse of the patent, change of ownership, withdrawal of the application, entrance into the national phase, etc. The records in this table originate from the patent gazettes and registers of various national patent authorities, including the EPO and WIPO. Currently over 50 offices provide the EPO with legal status data.

*European Patent Register database* – Released twice a year, the database contains bibliographic, legal and procedural information on published European patent applications and on published PCT applications for which the EPO is a designated office (so called Euro-PCT applications). The database is extracted from the European Patent Register, which stores all publicly available information the EPO has on European patent applications as they pass through the application and examination procedure. It includes information on: applicants, inventors, opponents and representatives, procedural events during application and examination proceedings, opposition and appeal proceedings, limitation and revocation proceedings.

*OECD REGPAT database* – It covers records on patent applications at the EPO (derived from Patstat) and PCT patents at international phase (derived from the EPO Bibliographic Database weekly downloads), for which addresses of inventors and applicants have been regionalised (*i.e.* assigned to a region code). The dataset covers regional information for most OECD and EU28 member countries, plus the BRICS countries. It can be linked to Patstat data using the *appln_id* field. All OECD databases are freely available on the OECD website. See Maraut et al. (2008) for technical details.

*OECD Triadic Patent Families (TPF) database* – It relies on a specific definition of patent family, covering patent applications filed at the EPO, the JPO and granted by the USPTO and that share a same set of priorities (Dernis and Khan 2004). The data is compiled using different patent linkages provided in Patstat and is a consolidated subset of the *tls219_inpadoc_fam* table. The *appln_id* field can be used to connect the data to Patstat.

*OECD Citations database* – It proposes a consolidated patent citation records of Patstat data for patents filed at the EPO or through the PCT. It mainly draws on the infrastructure proposed in Webb et al. (2005) and takes into account citations of patent and non-patent literature (NPL). In addition to the list of cited patents and NPL, it proposes a list of EPO or WIPO equivalents to patents cited, in order to facilitate further consolidation of the data.

*OECD Patent Quality Indicators database* – It proposes a number of indicators that are aimed at capturing the quality of patents and the possible impact that patent quality might have on subsequent technological developments, as described in Squicciarini et al. (2013). The current version of the dataset only relies on patent applications filed at the EPO but coverage will likely be expanded in the future to include patents filed to other offices. Indicators can be replicated using the program lines available in Squicciarini et al. (2013).

*OECD HAN database* – The OECD Harmonised Applicant Names (HAN) database proposes a grouping of patent applicant names resulting from a cleaning and matching of names.

*EEE-PPAT table* – In collaboration with the ECOOM department at KU Leuven, the EPO and Sogeti (a software consultancy), EUROSTAT has devoted considerable efforts to harmonize applicant names and allocate applicants to sectors (private business enterprises, universities and higher education institutions, governmental agencies, individuals). Sector allocation is relevant for analysing the constituents and dynamics of technological performance on the level of innovation systems. Read more at http://www.ecoom.be/en/EEE-PPAT

*EP-INV database on academic inventors* – It is the result of a project sponsored by the European Science Foundation and chaired by Francesco Lissoni. The database contains cleaned and standardised inventors' names and addresses, as well as information on the affiliations of academic scientists. See Den Besten et al. (2012) for more information.

*WIPO's IPC - Technology concordance table* – The WIPO technology concordance table links the International Patent Classification (IPC) symbols with thirty-five fields of technology. The concordance table is updated on a regular basis to reflect revisions to the IPC. Further information is provided on the WIPO website.

*Worldwide count of priority filings* – de Rassenfosse et al. (2013) have proposed an algorithm that exploits patent-family linkages (direct equivalents and other second filings) to recover missing information on inventor and applicant country of residence. Their algorithm can be used for the recovery of other information such as missing IPC codes.